\documentclass[preprint2]{aastex}
\usepackage{rotating}
\usepackage{graphicx}
\usepackage{grffile}

\begin{document}
\setcounter{secnumdepth}{2}
\title{Direct Spectrum of the Benchmark T dwarf HD~19467~B}
\author{Justin R. Crepp\altaffilmark{1}, Emily L. Rice\altaffilmark{2,5}, AAron Veicht\altaffilmark{5}, Laurent Pueyo\altaffilmark{3,4}, Jonathan Aguilar\altaffilmark{3,4}, Paige Giorla\altaffilmark{2,5,12}, Ricky Nilsson\altaffilmark{5,13}, Statia H. Cook\altaffilmark{5}, Rebecca Oppenheimer\altaffilmark{5}, Sasha Hinkley\altaffilmark{6,7}, Douglas Brenner\altaffilmark{5}, Gautam Vasisht\altaffilmark{8}, Eric Cady\altaffilmark{8}, Charles A. Beichman\altaffilmark{9}, Lynne A. Hillenbrand\altaffilmark{6}, Thomas Lockhart\altaffilmark{8}, Christopher T. Matthews\altaffilmark{1}, Lewis C. Roberts, Jr.\altaffilmark{8}, Anand Sivaramakrishnan\altaffilmark{3}, Remi Soummer\altaffilmark{3}, Chengxing Zhai\altaffilmark{8}}
\email{jcrepp@nd.edu} 
\altaffiltext{1}{Department of Physics, University of Notre Dame, 225 Nieuwland Science Hall, Notre Dame, IN, 46556, USA}
\altaffiltext{2}{College of Staten Island, CUNY, 2800 Victory Boulevard, Staten Island, NY 10314, USA}
\altaffiltext{3}{Space Telescope Science Institute, 3700 San Martin Drive, Baltimore, MD 21218, USA}
\altaffiltext{4}{Department of Physics and Astronomy, Johns Hopkins University, Baltimore, MD 21218, USA}
\altaffiltext{5}{American Museum of Natural History, Central Park West at 79th Street, New York, NY 10024, USA}
\altaffiltext{6}{Department of Astronomy, California Institute of Technology, 1200 E. California Blvd., Pasadena, CA 91125, USA}
\altaffiltext{7}{University of Exeter, Department of Physics and Astronomy, Prince of Wales Road, Exeter, Devon UK}
\altaffiltext{8}{Jet Propulsion Laboratory, California Institute of Technology, 4800 Oak Grove Drive, Pasadena, CA 91109, USA}
\altaffiltext{9}{NASA Exoplanet Science Institute, California Institute of Technology, Pasadena, CA 91125, USA}
\altaffiltext{10}{Caltech Optical Observatories, California Institute of Technology, 1200 E. California Blvd., Pasadena, CA 91125, USA}
\altaffiltext{11}{Institute of Astronomy, University of Cambridge, Madingley Road, Cambridge CB3 0HA, UK}
\altaffiltext{12}{Department of Physics, The Graduate Center, City University of New York, 365 5th Ave, New York, NY 10016}
\altaffiltext{13}{Department of Astronomy, Stockholm University, AlbaNova University Center, Roslagstullsbacken 21, 106 91 Stockholm, Sweden}

\begin{abstract}  
HD~19467~B is presently the only directly imaged T dwarf companion known to induce a measurable Doppler acceleration around a solar type star. We present spectroscopy measurements of this important benchmark object taken with the Project 1640 integral field unit at Palomar Observatory.  Our high-contrast $R\approx30$ observations obtained simultaneously across the $JH$ bands confirm the cold nature of the companion as reported from the discovery article and determine its spectral type for the first time. Fitting the measured spectral energy distribution to SpeX/IRTF T dwarf standards and synthetic spectra from BT-Settl atmospheric models, we find that HD~19467~B is a T5.5$\pm1$ dwarf with effective temperature $T_{\rm eff}=978^{+20}_{-43}$ K. Our observations reveal significant methane absorption affirming its substellar nature. HD~19467~B shows promise to become the first T dwarf that simultaneously reveals its mass, age, and metallicity independent from the spectrum of light that it emits.  
\end{abstract}                                                                                                                                                                                                                                                                                                                                                                                                                                                                                                                                                                                                                                                                                                                                                                                                                                                                                                                                                                                                                                        

\section{INTRODUCTION}\label{sec:intro}
HD~19467~B is a faint co-moving companion to the nearby G3V star HD~19467 that was recently discovered as part of the TrenDS high-contrast imaging program \citep{crepp_12b,crepp_14}. Prior to its direct imaging detection, the existence of HD~19467~B was first inferred from 16.9 years of precise stellar radial velocity (RV) measurements that revealed a long-term systemic acceleration of $dv/dt=-1.37\pm0.09$ m$\:$s$^{-1}\:$yr$^{-1}$. HD~19467~B's intrinsic brightness ($M_J=17.61\pm0.11$) and blue near-infrared colors ($J-H=-0.29\pm0.15$, $J-K_s=-0.36\pm0.14$) suggest that it is non-hydrogen-fusing. A lower mass limit of $M\geq51.9^{+3.6}_{-4.3}M_{\rm Jup}$ derived from the RV acceleration and projected orbital separation of $51.1\pm1.0$ AU is consistent with the model-dependent mass estimate, $M=56.7^{+4.6}_{-7.2}M_{\rm Jup}$, and supports the interpretation that HD~19467~B is a cold brown dwarf. \citealt{crepp_14} assigned a preliminary spectral-type of $\approx$T5--T7 based on the object's location in color-magnitude diagrams but spectroscopic observations have yet to be reported. 

HD~19467~B is an important benchmark object. It is currently the only directly imaged T dwarf companion known to cause a measurable Doppler acceleration around a nearby solar-type star. In addition to a precise parallax ($32.40\pm0.62$ mas) and many years of legacy RV measurements, astrometric observations already show systemic orbital motion at $22\pm6$ mas yr$^{-1}$ \citep{van_leeuwen_07}. A tight mass constraint from dynamics will be available once both the astrometry and RV observations reveal curvature \citep{crepp_12a}. Further, the chemical composition of HD~19467~B has been inferred from its Sun-like (G3V) parent star. \citealt{crepp_14} find that HD~19467~A has a sub-solar metallicity of [Fe/H]$=-0.15\pm0.04$, thus HD~19467~B should also have relatively low metal content assuming it formed from the same cloud of material. Such model independent mass and metallicity measurements are essential for calibrating theoretical spectral and evolutionary models, particularly at cold temperatures.

The age of the HD~19467 system is currently less certain but forthcoming observations will place strong constraints on the primary star's evolutionary state. At a distance of only $30.9\pm0.6$ pc, the interferometric CHARA array may be able to spatially resolve the surface of the star at near-infrared wavelengths, providing a direct measure of its radius \citep{boyajian_09}. It is expected that the age could be tightly constrained because HD~19467 resides $\Delta M_V=0.28$ above the Hipparcos median main-sequence, has a subsolar metallicity, yet is nearly identical to the Sun. Therefore, we already know that HD~19467 is older than the Sun ($> 4.6$ Gyr), and a radius constraint, even if only an upper-limit (i.e., marginally spatially resolved by CHARA interferometry), will eliminate ages older than some threshold value. 
%HD~19467's gyrochronological age is estimated empirically to be $4.3^{+1.0}_{-1.2}$ Gyr from its $\log{R'_{HK}}=-4.98\pm0.01$ and $B-V=0.65$ values \citep{crepp_14}. In this article, we show that theoretical evolutionary models for HD~19467~B suggest an age closer to 10 Gyr.

HD~19467~B has an ideal separation ($1.65\arcsec$) for follow-up integral field spectroscopy measurements (e.g., \citealt{bowler_10_hr8799b,konopacky_13,oppenheimer_13}). In this Letter, we use the Project 1640 high-contrast instrument at Palomar to measure HD~19467~B's spectral-energy distribution in the $JH$ bands and assess its physical properties. Spectral standard comparisons and theoretical model fits to the data allow us to assign a spectral type and directly study the object's atmosphere for the first time.  
 
\begin{figure*}[!t]
\begin{center}
\includegraphics[height=2.2in]{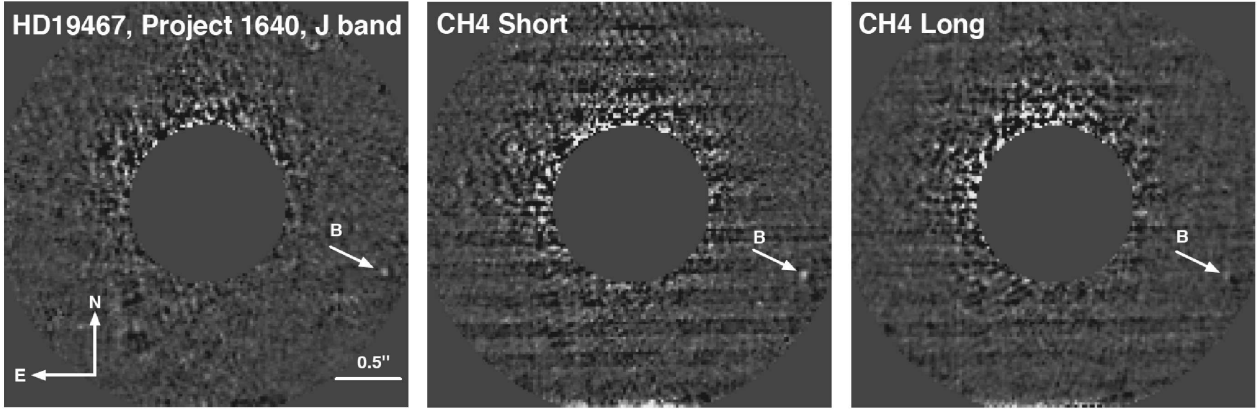} 
\caption{\label{images}Reconnaissance images of HD~19467~B taken with the Project 1640 integral field unit at Palomar Observatory using the 200-inch telescope on October 17, 2013. High-contrast measurements in multiple bands recover the companion following PSF subtraction. A marked decline in the flux of HD~19467~B in the $CH_4$ band $(\lambda_{\rm short}=1.560-1.600 \: \mu$m, $\lambda_{\rm long}=1.635-1.675 \: \mu$m) reveals significant methane absorption (see text for discussion). A description of $H$-band methane filter observations can be found in \citealt{janson_13}.} 
\end{center}\label{fig:image}
\end{figure*}  
 
\section{PALOMAR HIGH-CONTRAST OBSERVATIONS}
HD~19467 was observed on 2013 October 17 UT at Palomar Observatory using the Hale 200-inch telescope. The PALM-3000 AO system provides diffraction-limited images at near-infrared wavelengths using a Shack-Hartmann wavefront sensor \citep{dekany_13}. Given the brightness of HD~19467 ($V=7.0$), we used 64$\times$64 subapertures to provide fine spatial wavefront correction while running the AO system at 1 kHz. Simultaneous imaging and spectroscopy measurements were acquired by the Project 1640 (hereafter, P1640) high-contrast integral field spectrometer (IFS) \citep{hinkley_11_PASP}.

The seeing was estimated to be 1.3$\arcsec$ during the time of observations. We acquired 25 individual exposures totaling 4,583 seconds of on-source integration time with the star centered behind the coronagraphic mask. Unocculted images were acquired to measure the flux ratio between the primary and secondary as a function of wavelength. HD~19467~B resides just inside of the P1640 field of view ($4.2\arcsec \times 4.2\arcsec$). The companion can be noticed when playing a ``color movie" by sequentially viewing each image of the IFS data cube. Pre-speckle-suppressed images also show qualitatively that the companion appears to exhibit methane absorption in the H-band. We demonstrate that HD~19467~B must be a very cold object in the following analysis.

\subsection{Spectral Extraction}
Raw frames were converted into data cubes using the algorithm described in detail in \citealt{zimmerman_10}. Each exposure creates 32 separate images taken at slightly different wavelengths \citep{hinkley_11_PASP}. Although the companion light is initially mixed with stellar speckles, flux in each wavelength channel can be retrieved by performing PSF subtraction that takes advantage of color information provided by the IFS \citep{crepp_11,pueyo_12}. 

Once raw data frames were processed, we employed the techniques presented in Fergus et al. 2014 using the Spatial-Spectral model for Speckle Suppression (S4) algorithm. S4 uses principal components analysis (PCA) to identify a linear combination of orthogonal (KarhunenÐLoeve) modes to fit and remove the PSF of HD~19467~A in each wavelength channel (Fig. 1). A concise explanation for how S4 is adapted to P1640 can be found in Appendix B of \citealt{oppenheimer_13}. Various derivatives of PCA-based algorithms have been implemented previously for high-contrast imaging data (Amara \& Quanz 2012). We find that S4 provides consistent and stable behavior for extracting the spectrum of synthetic and real companions irrespective of relative brightness and angular separation for P1640 data cubes (Veicht et al. 2015, in prep.).  
% and is thoroughly discussed in Pueyo et al. 2014. S4 has also been applied using P1640 data for the substellar companion $\kappa$ Andromedae B \citep{hinkley_13}. 

HD~19467~B has an angular separation ($1.65\arcsec$) and $H$-band contrast ratio ($\Delta H=12.46\pm0.10$) comparable to the planet HR~8799~b \citep{marois_08}. Benefitting from correction of non-common-path errors from the P1640 internal wavefront calibration unit, our observations resulted in the companion being comparable in brightness to local speckles in frames spanning the $JH$-bands \citep{zhai_12,cady_13}. The relative flux of HD~19467~B was measured using aperture photometry with the occulted data cubes following speckle reduction by S4. Residual speckle noise was estimated by sampling many apertures at the same radial distance from the host star but at different azimuthal angles. 

The extracted companion flux was calibrated using several unocculted short exposures of the host star before the observing sequence to derive a spectral response function (SRF). The SRF was used to scale the final spectrum accordingly, thus accounting for wavelength-dependent sky and instrument transmission. We validated the SRF using a nearby G3V star (HD~10697) from the IRTF spectral library. Additional calibration sources (HIP~112821, HIP~43567) were also observed before and after HD~19467 to track airmass dependent and temporal variations in the SRF.  The high-resolution IRTF spectra were binned to match the wavelength sampling and coverage of the IFS. Comparing the SRF from HD~19467 to the above standard stars, we find that variations in the SRF are smaller than uncertainty in the companion flux introduced by the speckle suppression process. The fully calibrated spectrum of HD~19467~B is shown in Fig. 2. 

%The SRF used for calibrating HD~19467 B's spectrum was derived from HD~19467~A using unocculted frames. 
%To further validate the companion spectrum, we also processed the data cubes using a separate technique based on principle component analysis \citep{fergus_14}. Both approaches produce nearly indistinguishable results, i.e., well within measurement uncertainties in each spectral channel, so we proceed using the S4 reduction. For the brightest wavelengths, the uncertainty in this calibration dominates the spectro-photometric error budget??

\subsection{Astrometry}
The same speckle suppression procedures may be employed to self-consistently extract a precise location for the companion measured relative to the star. From S4 processed images we find that $HD~19467~B$ has an angular separation of $\rho=1.640\arcsec \pm 0.007\arcsec$ and position angle $PA=241.7^{\circ} \pm 0.3^{\circ}$. These values are consistent with that from \citealt{crepp_14} and confirm the measured systemic (clockwise) astrometric orbital motion. A longer time baseline however is required to constrain the companion mass and orbital parameters at levels much beyond that reported in the discovery paper.  

%$\Delta$RA   = - 1.4438'' $\pm$  0.0065"  \\
%$\Delta$DEC =  -0.7782''  $\pm$  0.0065"  \\

% PA=241.68^{\circ} \pm 0.34^{\circ}

\begin{figure*}[!t]
\begin{center}
\includegraphics[height=7.1in]{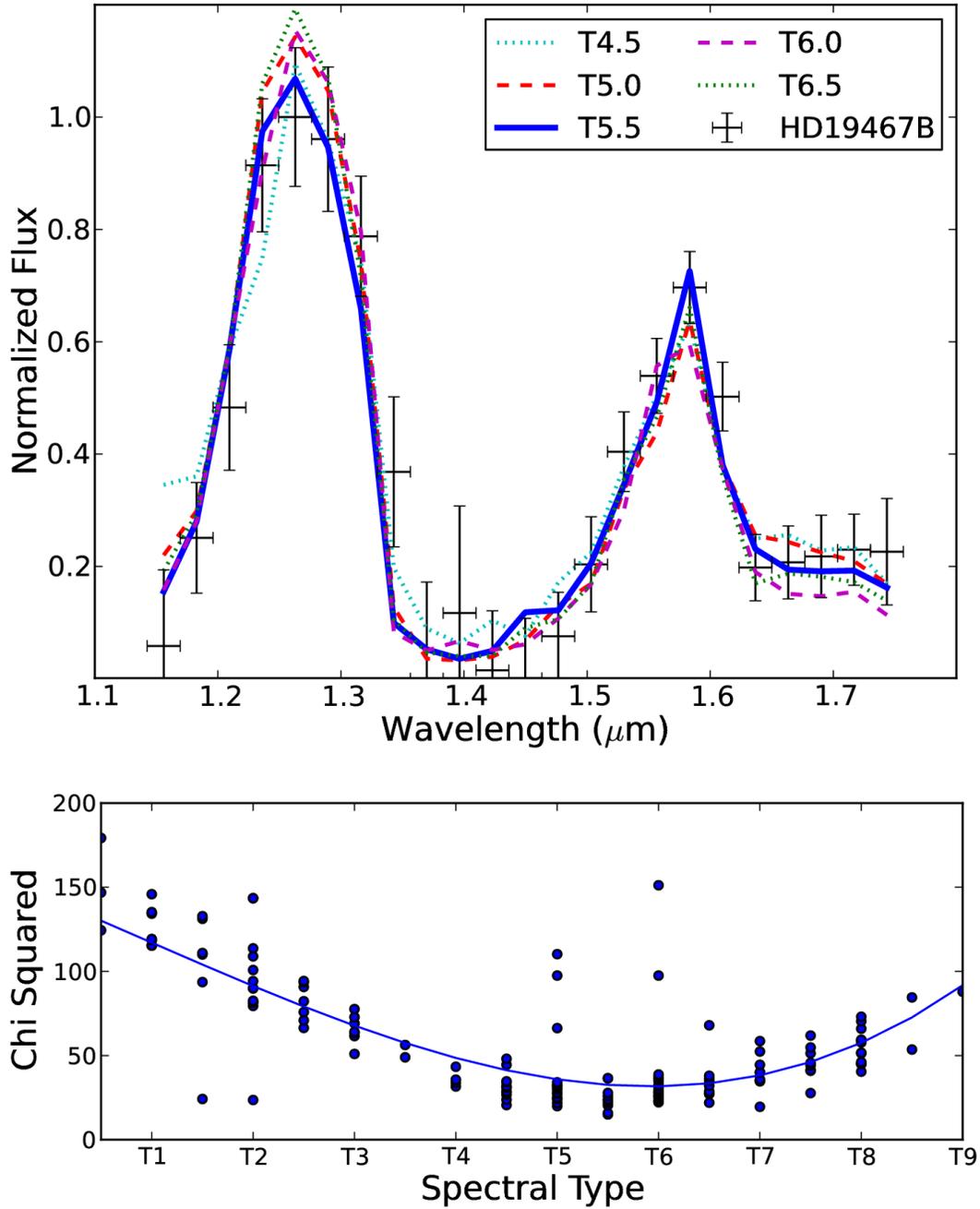} 
\caption{\label{spex_fit}(Top) P1640 spectrum of HD~19467~B (black crosses) plotted with binned and trimmed SpeX/IRTF spectra of T4.5--T6.5 objects. (Bottom) $\chi^2$ as a function of spectral type for T0--T9 objects and second-order polynomial fit. We derive a spectral type of T5.5$\pm$1 from this spectral comparison.} 
\end{center}\label{fig:image}
\end{figure*} 
% spectrum_fit_best_fit.eps
% spectral_type_chi2.eps

\begin{figure*}[!t]
\begin{center}
\includegraphics[height=7.0in]{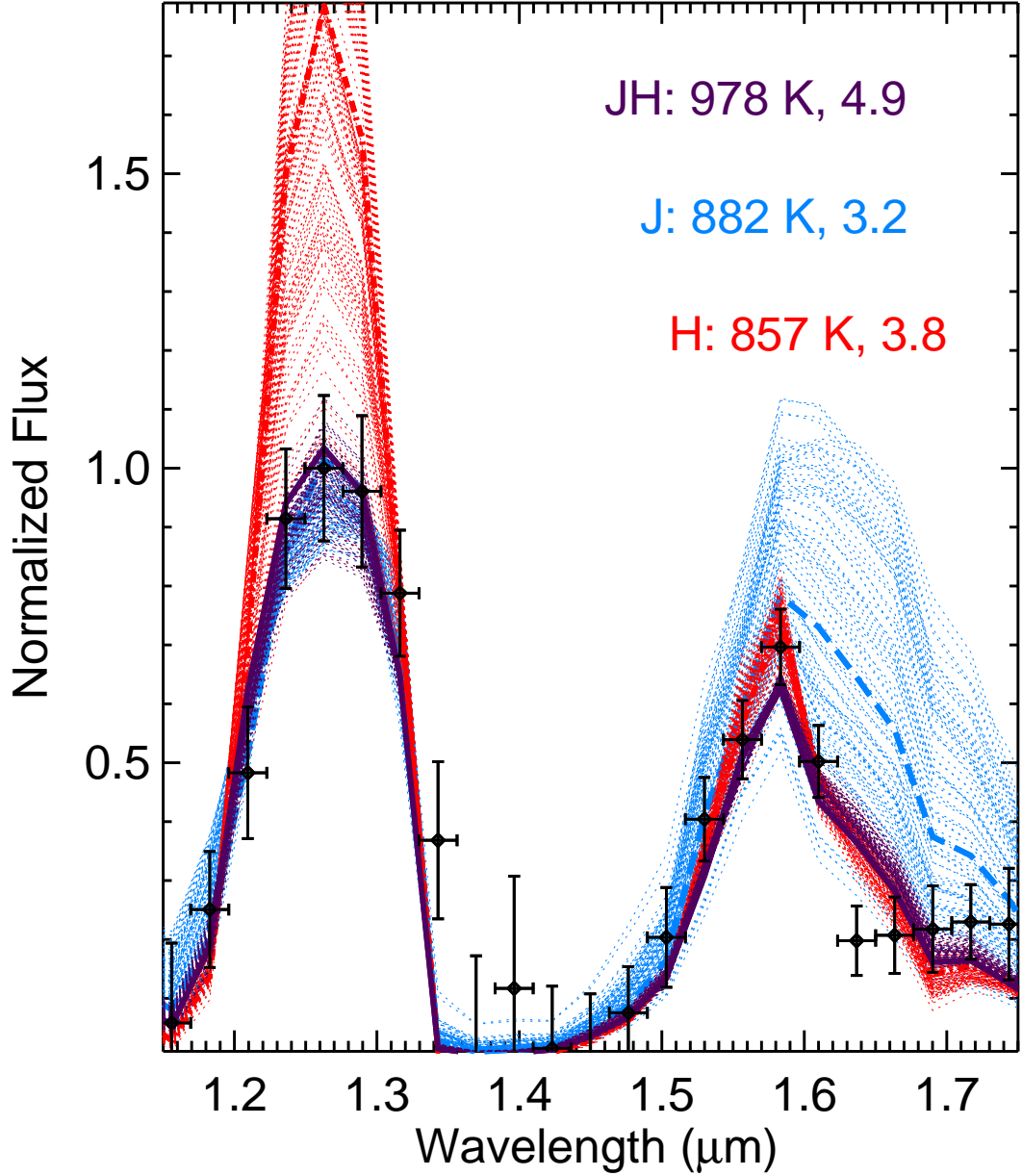} 
\caption{\label{BTSettl_fit}Best fit spectra from MCMC calculations for the complete $JH$ (purple), $J$-band (blue) and $H$-band (red) P1640 spectrum of HD~19467~B (black points and error bars). Thin dotted lines are 100 spectra randomly selected from the posterior distributions plotted in Figure~\ref{MCMC_pdfs} to represent the range of model fits allowed within 1$\sigma$ uncertainties. Values listed in the legend indicate the mode of the temperature and surface gravity from the posterior distribution. HD~19467~B has a best fit effective temperature of $T_{\rm eff}=978^{+20}_{-43}$ K.} 
\end{center}\label{fig:image}
\end{figure*} 
% HD19467B_bestfit_v2

\begin{figure*}
\includegraphics[height=6.8in]{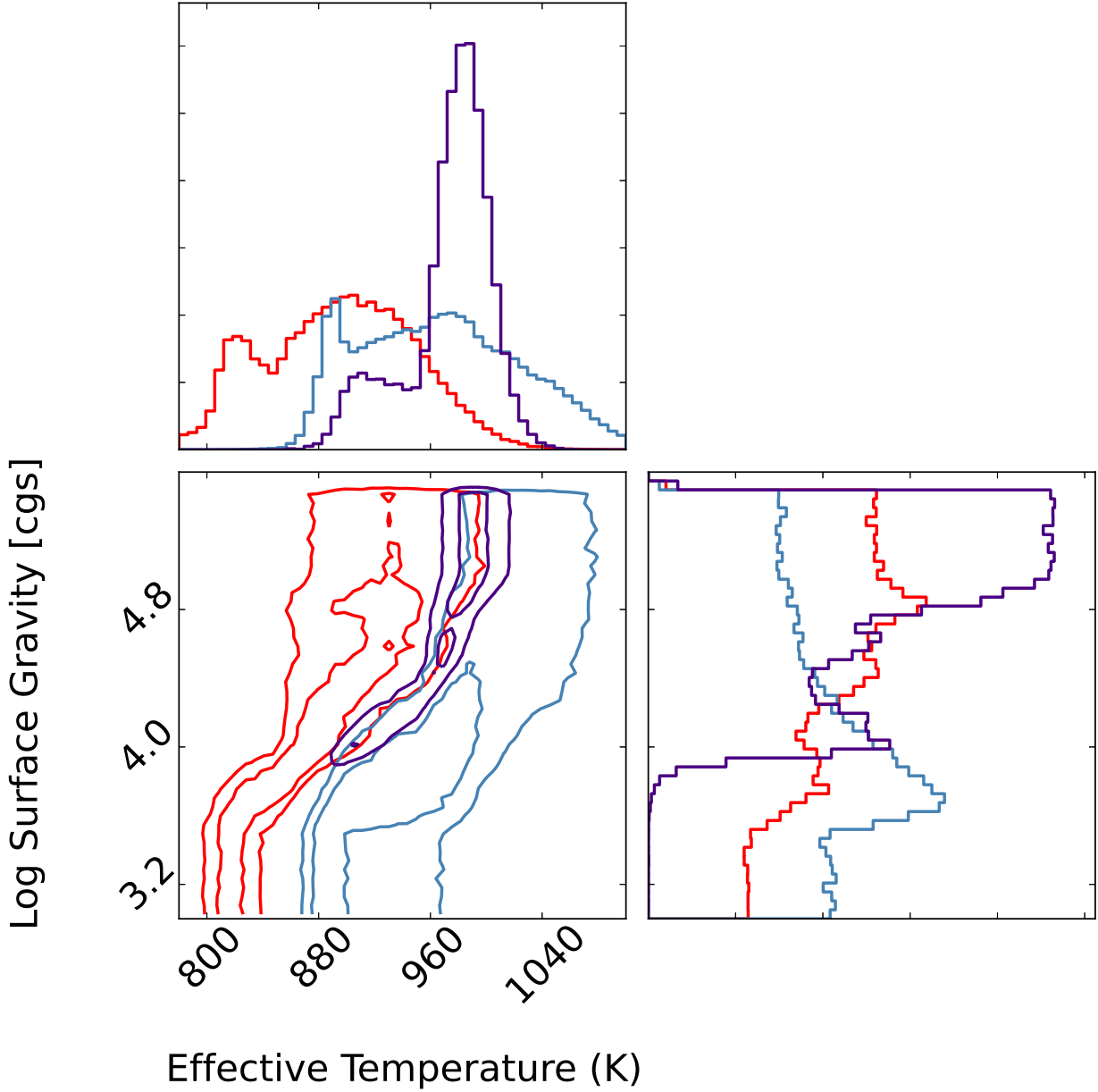}
\caption{\label{MCMC_pdfs}Posterior distributions using $10^6$ MCMC steps for $JH$-band (purple), $J$-band (blue), and $H$-band (red) spectra. Histograms show the distributions marginalized over gravity (top) and temperature (right). The distributions for different bands show multiple peaks indicating that higher spectral resolution measurements could further refine our understanding of the companion physical properties. HD~19467~B has a cool temperature, $T_{\rm eff}=978^{+20}_{-43}$ K, and surface gravity, $\log g =  4.87^{+0.44}_{-0.66}$ (see text for discussion).}
\end{figure*}
% HD19467B_JH_J_H

\begin{figure*}[!t]
\begin{center}
\includegraphics[height=7in]{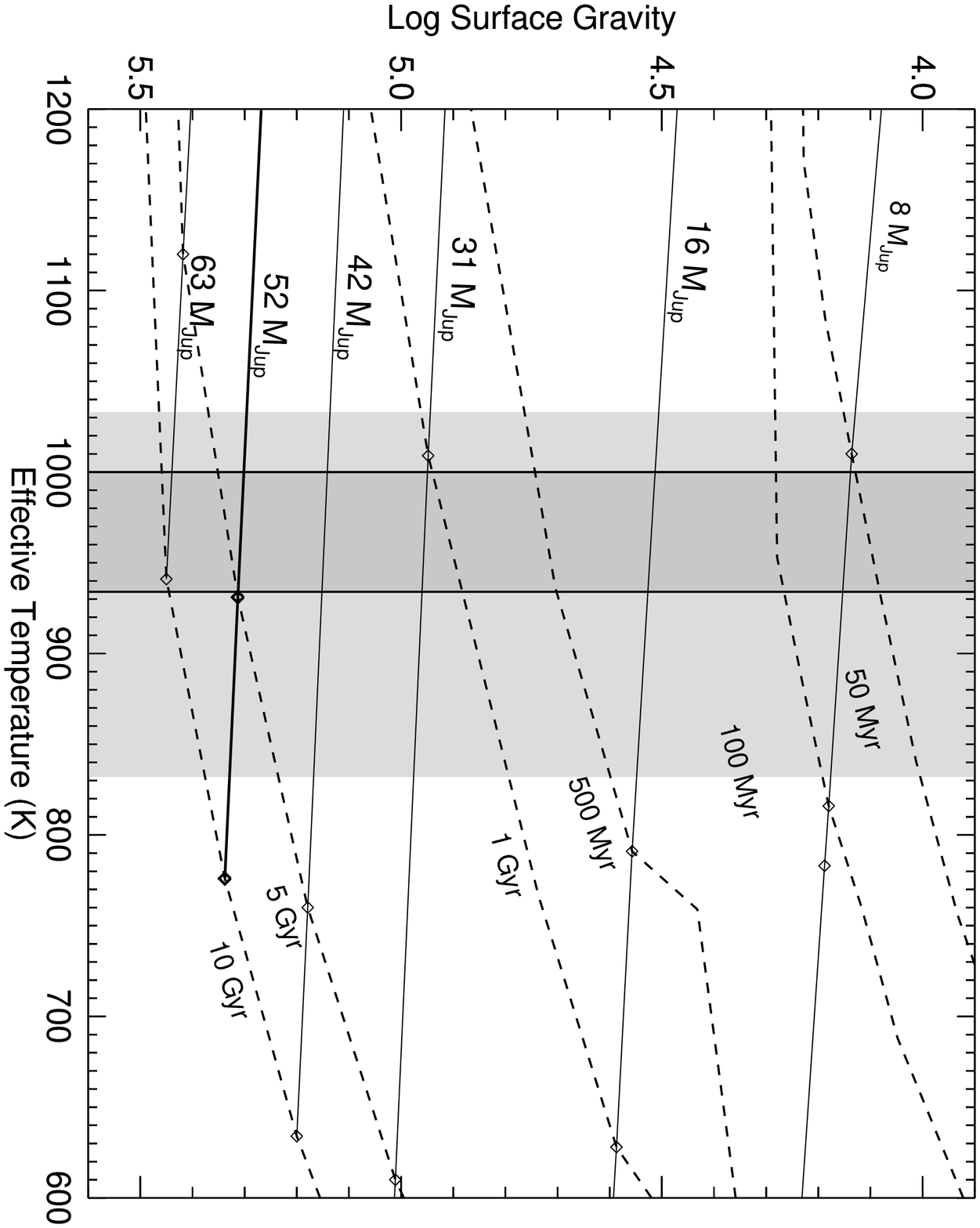} 
\caption{\label{COND03_tracks}Best-ft model parameters for surface gravity and effective temperature plotted onto COND03 isochrones and mass tracks (Baraffe the al. 2003). The dark gray shaded region shows the 1-$\sigma$ parameter space for the full $JH$ spectral fit discounting the (likely spurious) surface gravity results. The light-gray shaded region indicates the 1-$\sigma$ parameter space for $J$ and $H$ spectral fits. The effective temperature derived from spectral fits shows consistency with the 9$\pm$1~Gyr isochronal age of the primary star and $M>52M_{Jup}$ mass constraint (solid line) derived from the RV acceleration (Crepp et al. 2014). The $4.3^{+1.0}_{-1.2}$ Gyr gyro-chronological age is less consistent with the companion mass and effective temperature though cannot be ruled out.} 
\end{center}\label{fig:image}
\end{figure*} 

\section{SPECTRAL TYPING}
Crepp et al. (2014) estimate a spectral type of T5--T7 for HD~19467~B based on near-infrared broadband colors and $JHK$ absolute magnitudes \citep{leggett_10,dupuy_liu_12}. We refine this analysis using the $R=\lambda / \Delta \lambda \approx 30$ P1640 spectrum of HD~19467~B (Fig. 2). T dwarfs are optimally assigned a spectral type by comparing their near-infrared (0.8--2.5~$\mu$m) spectra to spectral standards (Burgasser et al. 2006a). We compare the P1640 spectrum of HD~19467~B to 107 T~dwarfs with spectral types ranging from T0 to T9. Spectra are from Burgasser et al. (2004, 2006a, 2008, 2010), Burgasser (2007), Chiu et al. (2006), Cruz et al. (2004), Kirkpatrick et al. (2011), Liebert \& Burgasser (2007), Looper, Kirkpatrick, \& Burgasser (2007), Mace et al. (2013), Mainzer et al. (2011), and Sheppard \& Cushing (2009). Most of the spectra were obtained via the SpeX prism library\footnote{\url{http://pono.ucsd.edu/~adam/browndwarfs/spexprism}}. 
%All such spectra were observed using the SpeX spectrograph (Rayner et al 2003) on the NASA IRTF in prism mode, producing 0.8--2.5 $\mu$m spectra with a resolving power $R \approx 120$. 

We first bin and trim the SpeX/IRTF spectra to match the wavelength range and resolution of the observations. We then calculate $\chi^2$ for each binned T dwarf spectrum compared to the measurements; $\chi^2$ is found by incorporating errors from both the P1640 spectrum of HD~19467~B and SpeX templates. A spectral type of T5.5 (2MASS J11101001+0116130, Burgasser et al. 2006) produces the minimum $\chi^2$ value slightly higher $\chi^2$ values for several other objects with spectral types T4 through T7. Figure~\ref{spex_fit} shows $\chi^2$ as a function of spectral type for T0--T9 dwarfs with a second-order polynomial fit. Spectral types earlier than T4 all result in higher $\chi^2$ values. We find a minimum (reduced) goodness-of-fit value of $\chi_r^2=0.68$ with 22 degrees of freedom. From this analysis, we derive a spectral type of T5.5$\pm$1, compatible with the estimated T5--T7 spectral type from \citealt{crepp_14} which was based on broadband photometry alone. 

Methane absorption represents another characteristic feature of cold brown dwarf atmospheres \citep{oppenheimer_95}. In fact, L and T dwarf spectral sequences may be distinguished based on the existence of $CH_4$ in the $H$-band. We estimate the $CH_4$ spectral index of HD~19467~B using the following metric from \citealt{geballe_02}:
\begin{equation}
S_{CH_4}=\frac{\int_{1.560}^{1.600} f_{\lambda} d\lambda }{\int_{1.635}^{1.675} f_{\lambda} d \lambda}
\end{equation}
where the integrals are carried out over the continuum (numerator) compared to wavelengths where $CH_4$ is present in late-type T dwarfs (denominator). Brown dwarfs have $S_{CH_4}>1$. The methane index increases monotonically for progressively lower temperature objects. Using $f_{\lambda}$ values from Fig. 2 we find that HD~19467~B has $S_{CH_4}=3.1^{+0.6}_{-0.8}$. These values correspond to an empirically determined spectral type of $\approx T6$ \citep{geballe_02}. While the Geballe metric is meant for interpretation of higher resolution spectra than that provided by P1640, we use it here as a qualitative indicator to demonstrate that HD~19467~B shows significant methane absorption and the spectral type found from this specific index is compatible with fits to the broader spectrum. 

% Previous methane index from KLIP: 2.91 {+0.54,-0.25}

\section{MODEL ATMOSPHERES}
T dwarf spectral types are defined by near-infrared spectral morphology and do not necessarily correspond monotonically to physical properties (Kirkpatrick et al. 2008). However, by comparing the observed spectrum to synthetic spectra from model atmospheres, we can infer parameters such as effective temperature and surface gravity. We compare the P1640 spectrum of HD~19467~B to synthetic spectra from the most recent BT-Settl model atmospheres (Allard 2014). The model grid was evaluated at increments of 50--100~K from 400 to 4500~K for surface gravities log(g)=3.5, 4.0, 4.5, 5.0, and 5.5 (cgs units). All models are solar metallicity. Our fitting procedure is summarized below. 

Model spectra are binned from their native resolution of $\Delta \lambda$=$1 \; \AA$ to match that of P1640 ($\Delta \lambda$ = $27$ nm). A goodness-of-fit parameter ($\chi^2$) is calculated for every binned spectrum in the model grid. We then generate probability distributions, $P\propto \exp{-\chi^2/2}$, using a 10$^6$-step Markov Chain Monte Carlo (MCMC) simulation with the Metropolis Hastings algorithm. The MCMC routine begins at the model with the smallest $\chi^2$ value and interpolates between model results. We find that jump sizes of 100--250~K in temperature and 1.0--2.5~dex in surface gravity produce optimal acceptance ratios of $\sim0.35$. 

We fit four versions of the extracted P1640 spectrum of HD~19467~B: the complete $JH$ spectrum, the $JH$ spectrum without the four flux points closest to the water absorption band at 1.4~$\mu$m (``trimmed" $JH$), and individual $J$-band and $H$-band spectra. Results for the complete versus trimmed $JH$ spectra are effectively identical so we present results for the complete $JH$ spectrum and the individual $J$- and $H$-band spectra (Fig. 3). We note that P1640 acquires $JH$ measurements simultaneously whereas most other high-contrast instruments must use separate $J$, $H$ filters for non-contemporaneous observations. We have decided to compare derived effective temperatures for individual filters to highlight the fact that results can vary appreciably depending on bandpass.  

Goodness-of-fit calculations ($\chi^2$) resulted in a global minimum between $700\leq T_{\rm eff} \leq 1100$~K at all surface gravities and using all four spectral fits. Posterior distributions in Figure~\ref{MCMC_pdfs} show multiple peaks in effective temperature with both the $J$-band ($902-1033$) and $H$-band ($832-949~K$) distributions producing slightly cooler results than the complete $JH$ data set (934--998~K). Using the mode posterior value and 68\% confidence interval from the $JH$ spectrum, we find that HD 19467 B has an effective temperature of $T_{\rm eff}=978^{+20}_{-43}$ K. This value is consistent with results for objects with similar spectral types from previous studies using bolometric luminosities (Golimowski et a. 2004, Vrba et al. 2004) and higher spectral resolution model comparisons (Burgasser et al. 2006b, del Burgo et al. 2009).

%%%%% NEWEST S4 RESULTS %%%%%
% JH band: 934 - 998, mode = 978, med=979;    log g = 4.21-5.31, mode=4.87, med=4.91
% J only: 902-1033, mode=882, med=966;          log g = 3.40-4.95, mode=3.21, med=4.03
% H only: 832-949, mode=897, med=896;           log g = 3.68-5.18, mode=3.83, med=4.53

%%%% FORMER KLIP RESULTS %%%%%
% J-band only: Teff = 731--890K
% H-band only: Teff = 711--808 K
% JH-bands: Teff = 731--843 K [median = 50% quantile = 790 K]

Posterior distributions for the surface gravity marginalized over effective temperature likewise show multiple peaks. The individual $J$-band and $H$-band results span 2 orders of magnitude but the combined $JH$ results show convergence towards higher surface gravity, $\log g = 4.21-5.31$. To test the veracity of the surface gravity results, we fit model spectra to 15 T dwarfs from the SpeX Prism library binned and trimmed to match P1640. The template objects include nine spectral standards (T0 to T8), three additional late-type objects (T7.5, T8, and T8pec) and two early T dwarfs, including the young $\approx0.3$~Gyr companion HN~Peg~B (Luhman et al. 2007). We find that the most recent BT-Settl models reproduce the $R\approx30$ spectra for cold dwarfs later than T5. All of these objects are $>1$ Gyr old with surfaces gravities predicted to be $>$5.0~dex (e.g., Baraffe et al. 2003), with the exception of HN~Peg~B. Model fits for field objects later than T5 result in best fit surface gravities  $\log g=4.1-4.9$~dex, while the best fit surface gravity for HN~Peg~B, which is known to be young, is $>$5.0~dex. Therefore we conclude that our surface gravity results are not reliable, likely a combination of the low spectral resolution perhaps exacerbated by a systematic bias in the BT~Settl synthetic spectra.

\section{EVOLUTIONARY MODELS}
Initial analysis of the HD~19467 system age from \citealt{crepp_14} yielded two distinct results: a gyro-chronological age of $4.3^{+1.0}_{-1.2}$ Gyr, and an isochronal age of $9\pm1$ Gyr. The younger age was adopted in that paper due to convergence issues when iterating the ``Spectroscopy Made Easy" program for tracks near the end of the stellar evolutionary model grid \citep{valenti_fischer_05}. Given the effective temperature range that we find for HD~19467~B, we can now help discriminate between the above disparate age estimates. Figure~\ref{COND03_tracks} shows the best-ft model parameters for surface gravity and effective temperature over-plotted against COND03 isochrones and mass tracks \citep{baraffe_03}. The gray shaded region indicates the allowable parameter space for $T_{\rm eff}$ discounting the (possibly spurious) surface gravity results. The dark solid line corresponds to the mass lower-limit of $M>52M_{Jup}$ derived for HD~19467~B from RV measurements.  

We find that the gyrochronological age of $4.3^{+1.0}_{-1.2}$ Gyr corresponds to masses that lie near the edge of acceptable values ($m>52M_{Jup}$) for the range of temperatures allowed by spectral fitting. The companion would require an edge-on orbit in this case. Examining the intersection of the hottest allowable temperature, i.e., $1\sigma$ ($998$ K), from $JH$ spectral fitting and smallest allowable mass track ($M=52M_{Jup}$), we find that ages $\geq 4.6$ Gyr satisfy all available constraints \citep{baraffe_03}. This result favors the $9\pm1$ Gyr isochronal age, which was derived in \citealt{crepp_14} from color magnitude diagrams and high resolution stellar spectroscopy, but does not rule out the gyro-chronological age. In the absence of any additional diagnostics, we adopt the 4.6-10 Gyr range as the new current best estimate for the age of HD~19467. While it is not clear why the estimated gyrochronological age differs significantly from the isochronal age, it is perhaps not surprising to find a slightly evolved G3V stars with sub-solar metallicity that is older than the Sun. As discussed in the introduction, the age of HD~19467 will be subject to further testing with interferometry by directly measuring the radius of the primary star. 

% Self-consistent solutions are however available in the mass-T$_{\rm eff}$ parameter space for the $9\pm1$ Gyr isochronal age estimate of the primary star.
%It is not unreasonable to expect ages that reside in between the above estimates. 
% and/or are too cool for dynamical mass of the object predicted by the COND03 evolutionary models.

\section{SUMMARY \& CONCLUSIONS}
We present the first direct spectrum of currently the only T dwarf companion known to produce a Doppler acceleration around a solar type star. Our observations affirm the cold nature of the companion reported in the discovery article. By self-consistently comparing IRTF SpeX observations of cold brown dwarfs to low-resolution $JH$ spectra obtained by the P1640 IFS at Palomar, we find that HD~19467~B has an effective temperature of $T_{\rm eff}=978^{+20}_{-43}$ K. The spectral-type originally assigned by \citealt{crepp_14} is confirmed; our $R\approx30$ observations indicate that only a marginal refinement from $\approx$T5--T7 to T5.5$\pm1$ is required.   

Several lines of evidence indicate that HD~19467~B is a substellar object. P1640 spectra reveal significant methane absorption in the $H$-band. At $R\approx30$, we measure a spectral index of $S_{CH_4}=3.1^{+0.6}_{-0.8}$, consistent with that of a late T dwarf. At a temperature of $978$ K, evolutionary models predict that such an object must be substellar at any age \citep{baraffe_03}. Further, the dynamical mass limit derived from joint RV and imaging measurements yields a value of $M\geq52M_{\rm Jup}$, well within the brown dwarf range. 

Comparing the $T_{\rm eff}$ of HD~19467~B to evolutionary models resulted in a necessary revision to the companion age from $4.3^{+1.0}_{-1.2}$ Gyr (gyrochronological estimate) to $4.6-10$ Gyr. This is the only way to establish a self-consistent connection between our new spectroscopy results and the companion mass lower limit derived from the Doppler trend. 

%It is worth reiterating that none of the BT~Settl theoretical spectral models are consistent with the companion surface gravity and temperature. 

HD~19467~B presents a rare opportunity to study a cold brown dwarf for which mass, age, and metallicity information is available simultaneously in addition to a direct spectrum. Other comparable objects include GJ~758~B \citep{thalmann_09,janson_11} and GJ~504~b \citep{kuzuhara_13}. These late-type companions orbit nearby, well-characterized G-stars, are older than directly imaged exoplanets, and exhibit methane absorption \citep{janson_11,janson_13}, although do not yet have RV accelerations or spectra reported in the literature. Together, they may ultimately be used as a control group to study the effects of surface gravity on the atmospheres of the oldest and coldest benchmark TY-dwarfs. 

Theoretical spectral models and evolutionary models have yet to be calibrated over a wide range of temperatures and surface gravities. Indeed, the few benchmark objects that have been studied in sufficient detail to test these models have already revealed discrepancies in the luminosity values predicted from brown dwarf evolution \citep{dupuy_14}. Unlike T dwarfs discovered as field objects, the physical properties of HD~19467~B may be measured without relying upon theoretical models. Our current best estimates of the companion mass, age, and metallicity are: \\ \\
Mass: $\hspace{0.32in} \geq51.9^{+3.6}_{-4.3}M_{\rm Jup}$ (dynamics), \\
$\mbox{[Fe/H]}$: $\hspace{0.24in} -0.15\pm0.04$  \hspace{0.14in} (stellar spectroscopy), \\ 
Age: $\hspace{0.40in} 4.6-10$ Gyr \hspace{0.07in} (multiple techniques). \\ \\
respectively. The first two quantities have been determined without any reference to the emergent spectrum of light received from the companion itself. Only the age lower-limit has been inferred using brown dwarf theoretical evolutionary models. The age upper-limit was determined through stellar isochrones. In time, the uncertainty in each will narrow through continued Doppler and astrometric monitoring as well as further investigations of its host star, which has a precise parallax and conveniently resembles the Sun. 

\section{ACKNOWLEDGEMENTS}
The TrenDS high-contrast imaging program is supported by NASA Origins of Solar Systems grant NNX13AB03G and the NASA Early Career Fellowship program. A portion of this work was supported by the National Science Foundation under Grant Numbers AST-0215793, 0334916, 0520822, 0804417 and 1245018. This work was partially supported by NASA ADAP grant 11-ADAP11-0169 and NSF award AST Ð1211568. A portion of the research in this paper was carried out at the Jet Propulsion Laboratory, California Institute of Technology, under a contract with the National Aeronautics and Space Administration. JA is supported by the National Physical Science Consortium. This research has benefitted from the SpeX Prism Spectral Libraries, maintained by Adam Burgasser.\footnote{http://pono.ucsd.edu/~adam/browndwarfs/spexprism}

\end{document}